# Electromagnetic and weak processes with light halo nuclei

**E.M. Tursunov [1,2], D. Baye [1, 3], P. Descouvemont [2]**

[1] Physique Nucléaire Théorique et Physique Mathématique, C.P. 229,

Université Libre de Bruxelles, B 1050 Brussels, Belgium

[2] Institute of Nuclear Physics, Uzbekistan Academy of Sciences, 100214, Ulugbek, Tashkent, Uzbekistan

[3] Physique Quantique, C.P. 165/82,

Université Libre de Bruxelles, B 1050 Brussels, Belgium

## Abstract

The $\beta$-decay of the $^6$He halo nucleus and the M1-transition processes of the excited $^6$Li($0^+$) state into the $\alpha$ + d continuum are studied in a three-body model. The initial nuclear states are described as an $\alpha$ + 2N system in hyperspherical coordinates on a Lagrange mesh. The energy dependence and absolute values of the $\beta$-transition probability per time and energy units of a recent experiment can be reproduced very well with an appropriate $\alpha$ + d potential. A total transition probability of $2.04\times10^{-6}$ s$^{-1}$ is obtained in agreement with the experiment. Due to a strong cancellation of the internal and halo components of the matrix elements, halo effects are shown to be very important. For the M1 transition process the halo structure of the excited $^6$Li($0^+$) isobaric analog state is confirmed. The cancellation effects, however, are not so strong. The transition probability is strongly sensitive to the description of the $\alpha$ + d phase shifts. Charge symmetry is analyzed through a comparison of the two processes. The present branching ratio $\Gamma_\gamma(0^+ \rightarrow \alpha+d)/\Gamma_\gamma(0^+ \rightarrow 1^+)$ is of the order $10^{-4}$, and is expected to be observable in current experiments.

## 1. Introduction

A new interest in nuclear physics began with the discovery of light halo nuclei [1] with very large matter radii near the neutron drip line. The large radii are interpreted as arising from an extended spatial density of a few neutrons [2,3]. Of special interest is the study of the electromagnetic and weak processes in light halo nuclei. The beta decay with emission of a deuteron (beta-delayed deuteron decay) is energetically possible for nuclei with a two-neutron separation energy $S_{2n}$ smaller than 3 MeV. It was first observed for the $^6$He halo nucleus (see [4-6]). The difficulty of the measurement led to conflicting results and raised a number of questions. The problem was connected with the experimental branching ratio much smaller than expected from simple the R-matrix method [4], and from two-body [7] and 3-body [8] models.

A semi-microscopic study [9] of the process has explained that the low value of the branching ratio is the result of a cancellation between the "internal" and "external" parts of the Gamow-Teller matrix elements. The overlaps of the $^6$He ground state and $\alpha$ + d scattering wave functions in the internal ($R<5$ fm) and external ($R>5$ fm) regions have very close magnitudes but opposite signs. It is clear that the external part of the Gamow-Teller matrix element reflects the properties of the halo structure of the $^6$He nucleus. This picture of the beta-delayed deuteron decay of the $^6$He nucleus was confirmed in microscopic [10,11] models and in the R-matrix framework [12]. The model [11] provided a reasonable agreement with the data of [5]. Without any fitted parameter, those data were underestimated by about a factor of 2. Hence, the same microscopic results now overestimate the recent data of [6] by a similar factor.

It is also very interesting to study the M1-transition of the isobaric analog state $^6$Li(0$^+$,T=1) of the $^6$He halo nucleus into the $\alpha$ + d continuum. The branching ratio between the transition probabilities to the continuum and to the $^6$Li(1$^+$) ground state was estimated as $8\times 10^{-5}$ with a number of assumptions [13]. However the shape and magnitude of the transition probability are much more informative than a simple number. These studies are important to test the halo-structure of the excited $^6$Li(0$^+$) state in comparison with the $^6$He halo nucleus. The gamma-delayed deuteron spectrum of $^6$Li(0$^+$) should test charge symmetry of the strong interaction in exotic light nuclei.

The aim of the present work is the determination of the deuteron spectrum shape and branching ratio for the beta decay of the $^6$He halo nucleus and for the M1 transition of the excited $^6$Li(0$^+$) state into the $\alpha$ + d continuum. We base our discussion on an $\alpha + N + N$ three-body model. Very accurate wave functions of $^6$He and $^6$Li nuclei are available in hyperspherical coordinates [14]. For the description of the structure of the $^6$He and $^6$Li nuclei, we use the hyperspherical harmonics method on a Lagrange mesh [14,15] which yields an accurate solution in a three-body cluster model. The $\alpha$ + d scattering wave function is treated as factorized into a deuteron wave function and a nucleus-nucleus scattering state. We will choose several versions of the central interaction potential: a deep Gaussian potential [16] which fits both the *S*-wave phase shift and the binding energy of the $^6$Li ground state (1.473 MeV), and potentials obtained by folding the $\alpha + N$ potential of Voronchev et al. [17]. For the sake of comparison we will also perform a calculation with a repulsive $\alpha$ + d potential which was used by the authors of Ref. [8].

## 2. Model

For the $\beta$-decay process

$$^{6}\mathrm{He} \to \alpha + \mathrm{d} + e^{-} + \overline{\upsilon}, \tag{1}$$

the transition probability per time and energy units is given by

$$\frac{dW}{dE} = \frac{m_e c^2}{\pi^4 v \hbar^2} G_\beta^2 f(Q-E) B_{GT}(E), \tag{2}$$

where $m_e$ is the electron mass, $v$ and $E$ are the relative velocity and energy in the center of mass system of $\alpha$ and deuteron, $G_\beta$=2.996 × $10^{-12}$ is the dimensionless $\beta$-decay constant [18]. The Fermi integral $f(Q-E)$ depends on the kinetic energy $Q$-$E$ available for the electron and the antineutrino. The mass difference between the initial and final particles is 2.03 MeV.

Since the total orbital momentum and parity are conserved, only the $l=0$ partial scattering wave contributes. Hence, only the initial $L=S=0$ component of the $^6$He can decay to the $\alpha$ + d continuum. In this case the non-zero matrix elements are obtained for the three-body components with $l_x = l_y = L = S = 0$ :

$$Z_K(r,R) = [(A-2)/A]^{3/4} \rho^{-5/2} \chi_{0000K}(\rho) N_K P_{K/2}^{1/2,1/2}(\cos 2\alpha), \tag{3}$$

where $N_K$ is a normalization factor, $\chi(\rho)$ an hyperradial wave function and $P_{K/2}^{1/2,1/2}(x)$ a Jacobi polynomial. After introducing the effective functions

$$u_{eff}^{(K)}(R) = R\int_0^\infty Z_K(r,R) u_d(r) r dr, \tag{4}$$

and the effective integrals

$$I_E^{(K)}(R) = \int_0^R u_E(R') u_{eff}^{(K)}(R') dR' \tag{5}$$

for the reduced Gamov-Teller transition probability, one can write

$$B_{GT}(E) = 6\lambda^2 [\sum_K I_E^{(K)}(\infty)]^2, \tag{6}$$

which relates the transition probability to the values of the effective integrals at infinity.

For the M1 process

$$^{6}\mathrm{Li}(0^{+}) \rightarrow \alpha + \mathrm{d} + \gamma, \tag{7}$$

the transition probability per time and energy units is related to the reduced matrix elements of the M1 operator between the initial $^{6}$Li(0$^{+}$) and final $\alpha$ +d states as

$$\frac{dW_{\gamma}}{dE} = \frac{32\mu m_N k^3}{3\hbar^3 k_d} \left\langle \Psi_{\alpha d}(1^{+}) \| M1 \| \Psi_{6Li}(0^{+}) \right\rangle^2, \tag{8}$$

where the dimensionless reduced mass is $\mu$ = 4/3, and $k$ and $k_d$ are the wave numbers of the emitted photon and of the $\alpha$+d relative motion, respectively (see [19] for details). We note that the transition probability is also directly proportional to the same effective integrals (5) as in the case of the $\beta$-decay process of the $^{6}$He nucleus into the $\alpha$ +d continuum.

Thus for both processes (1) and (7) it is very useful to investigate the convergence of the results based on the effective integrals. This allows us to test additionally the charge symmetry properties of the nuclear forces.

## 3. Conditions of the calculation

The initial α+n+n bound state is calculated as explained in Ref. [14]. The same nucleon-nucleon interaction is used, i.e., the central Minnesota interaction [20] which reproduces the deuteron binding energy and fairly approximates the low-energy nucleon-nucleon scattering. This potential provides the deuteron wave function $\Psi_d$. The α + *n* potential is however different from the one employed in Ref. [14]. Here we employ the *α + n* potential of Voronchev et al. [17] with a multiplicative factor 1.035 in order to reproduce the $^{6}$He binding energy. This change of interaction is motivated by the fact that we want to use the same interaction for the derivation of the *α* + d folding potential.

Since the valence neutron and proton in the $^{6}$Li nucleus belong to the $0p_{3/2}$ spectrum, we use the *P*-wave *α* + *N* potential of Ref. [17] when deriving the *α*+ d folding potential. The *S*-wave *α* + d folding potential derived from the *α +N* potential yields two bound states for $^{6}$Li with $E_0$ = −19.87 MeV and $E_1$ = −0.83 MeV, respectively. The first one is forbidden by the Pauli principle and the second one is underbound compared with the experimental ground-state energy $E_{exp}$ = −1.473 MeV. The *α* + N potential of Kanada et al. [21] employed in Ref. [14] does not yield an *α +* d folding potential with a physical bound state in the *S*-wave.

The numerical calculations of the Gamow-Teller matrix elements for the β decay of the $^{6}$He nucleus are done with *α +* d folding potentials, with a phenomenological Gaussian attractive potential [16], and with a Woods-Saxon repulsive potential $V_r$ [8]. The simple Gaussian potential $V_a$ simultaneously provides the correct $^6$Li binding energy (together with a forbidden state) and a fair fit of the low-energy experimental phase shifts. Since the folding potential does not reproduce the $^6$Li ground-state energy, we multiply the central part of the original $\alpha + N$ potential by factor $f_a$= 1.068. The corresponding folding $\alpha$ + d potential $V_{f1}$ puts the physical state at $E_1$= −1.470 MeV. The *S*-wave phase shifts for the different $\alpha + d$ potentials are compared in Fig. 1 with phase-shift analyses [22, 23]. The folding potential $V_{f1}$ does not have the same quality of phase shift description as the simple Gaussian potential. Therefore, we also choose another factor $f_b$ = 1.15 for the folding potential $V_{f2}$, which gives a stronger binding for the $^6$Li ground state, $E_1$= −2.386 MeV. From Fig. 1, one can see that the description of the $\alpha$ + d phase shift is poor for the Woods-Saxon repulsive potential, which does not bind the $^6$Li ground state.

## 4. Effective integrals

In Fig. 2, we display the integrals $I_E^{(K)}(R)$ at the $\alpha$ + d relative energy $E$ = 1 MeV for different $K$ values. They are obtained by using the $\alpha$ + d potential $V_a$ of Ref. [16]. From Fig. 2, one can see that at large $R$ values the dominant contribution to $I_E^{(K)}(R)$ for all $K$ values up to $K_{max}$ comes from the $K$ = 2 and $K$ = 8 components. Although the $K$ = 0 component is rather important around $R$ = 5 fm, it is suppressed at large $R$ values even more than the $K$ = 10 component.

One can observe (see [19]) that $u_{eff}^{(0)}(R)$ keeps a constant sign over the whole region while $u_{eff}^{(2)}(R)$ changes sign at $R \approx 2$ fm. The $\alpha$+d *S*- wave scattering function for $E$ = 1 MeV keeps a constant sign in the interval 5-19 fm. This constant-sign interval is even broader for smaller values of $E$. One can see that the product $u_{eff}^{(K)}(R)u_E(R)$ for $K = 0$ changes sign several times. The integral $I_E^{(0)}$ is first positive, starts decreasing at the first zero of $u_E$, changes sign near 2 fm and increases again at the second zero of $u_E$. The combined effect of both zeros results in a cancellation between the internal and external parts of the corresponding integral $I_E^{(0)}$. These zeros at short distances are due to the existence of two (one physical and one forbidden) bound states in the potential. The cancellation would not occur so strongly with a single zero. The combined effects of the zero of $u_{eff}^{(2)}(R)$ and of the fist zero of $u_E(R)$ is just a small plateau near 2 fm. The second zero of $u_E(R)$ gives a minimum near 5 fm. The result for the $K=2$ component also yields an important cancellation, but not as strong as in the $K=0$ case. The effective functions for $K = 4$ and $6$ are very small in the region where $u_E$ keeps a constant sign and lead to negligible contributions.

A new situation appears for the $K=8$ component. The effective wave function is much smaller than $K=0$ or $2$ but the cancellation is minimal. Hence it gives the second largest $I_E^{(K)}$ at infinity. The same mechanism applies for the smaller *K=10, 12, 14*, ... components. The *K=10* integral still contributes significantly to the total sum.

The strong cancellation found for the attractive potential $V_a$ does not occur for other potential models: folding potentials $V_{f1}$ and $V_{f2}$ and a repulsive Woods-Saxon potential $V_r$ . The repulsive potential displays a strongly different behavior from the other potentials. Thus, we find that the *K = 2* and *K = 8* components of the three-body hyperspherical wave function of the $^6$He nucleus give dominant contributions to the integral $I_E(R)$ at large *R* values and to the Gamow-Teller reduced transition probability $B_{GT}(E)$. This finding contradicts the statement of the authors of Ref. [8], that the contributions of the *K=0* and 2 components are dominant. The convergence with respect to the value of maximal hypermomentum is reached at $K_{max}$= 22 (see [19] for details).

At the next step we check the convergence behavior of the effective integrals for the M1 transition of $^6$Li(0$^+$) into the $\alpha$ + d continuum . In Fig. 3 we display the integrals calculated at *E*=1 MeV with the above $\alpha$ + d potentials and with the phase-equivalent potential $V_a^{S1}$ which has exactly the same $^6$Li ground-state energy and the same *S*-wave phase shift as $V_a$ but its scattering wave functions have one node less at small distances. It is obtained by using a pair of supersymmetric transformations [24]. Additionally we show here the effective integral from the beta decay of the $^6$He calculated with the $V_a$ potential model. One can see from this picture that the cancellation effects are not so strong as the case of the beta-decay. The dominant contribution to the process comes from the *K*=0, *K*=2 and *K*=8 components of the three-body hyperspherical wave function. Another point observed from this figure is that, in the internal region (*R*< 5 fm), the integrals for the both processes are close to each other, however they are different at large distances. In this way one can conclude about the charge symmetry property of the strong nuclear forces.

## 5. Delayed beta-decay transition probabilities

In Fig. 4, we display the transition probability for different potential models with the same $K_{max}$=24, and $R_{max}$=30 fm. One can see that more or less good description of experimental data is obtained with the attractive Gaussian potential $V_a$. It slightly underestimates the experimental data, but reproduces the shape of their energy dependence The worst results correspond to the repulsive Wood-Saxon potential, which does not give any bound state for $^6$Li and for which the description of the *S*-wave phase shift at low energies is poor. The folding potentials $V_{f1}$ and $V_{f2}$ have intermediate behaviors. Potential $V_{f1}$ overestimates the recent data significantly while potential $V_{f2}$ provides a better order of magnitude but its energy dependence disagrees with the experimental one. A relative success of the deep Gaussian potential could be attributed to the fact that it simultaneously reproduces both the $^6$Li ground state binding energy and the *S*-wave phase shift at low energies. However the discussion of Figs. 2-3 indicates that an important ingredient is the existence of two nodes in $u_{eff}$. Therefore it is natural to test the potential $V_a^{S1}$ for the beta-transition process. The resulting $dW/dE$ is about one order of magnitude larger and resembles the one obtained with the folding potential $V_{f1}$ (see Fig. 4). Notice however that $V_{f1}$ has two bound states but the phase shifts are not well reproduced. A second phase-equivalent potential $V_a^{S2}$ is obtained by removing the $^6$Li ground state from $V_a^{S1}$ with another pair of transformations. This repulsive potential has still exactly the same phase shifts as $V_a$ but no bound state. Its scattering wave functions have no node near the origin. The corresponding transition probability $dW/dE$ is now very close to the one obtained with potential $V_r$. The comparison emphasizes the crucial role played by the forbidden bound state, in addition to the physical $^6$Li ground state, for reproducing the order of magnitude of the experimental data.

In Fig. 4 we also show results with the modified $\alpha$ + d potential model $V_m(r)$ = -79.4 exp(-0.21 $r^2$) which slightly differs from the potential $V_a$ and reproduces both the phase shifts and the ground state energy value with two additional bound states. The resulting transition probability is very close to the new data from Ref. [6].

The total transition probabilities for different potentials are given in Table I. The second row contains results corresponding to the experimental cutoff [6]. The values in the last column are derived from the most recent experimental branching ratios and from the $^6$He half life [6]. As expected from the previous discussion, the result obtained with the Gaussian potential $V_M$ falls within the experimental error bars. The other results are too large, especially with the repulsive potential. We have also calculated the Gamow-Teller matrix elements for the β decay to the $^6$Li ground state. The value $B_{GT}$ = 4.489 $\lambda^2$ obtained with a three-body $^6$Li wave function calculated under the same conditions and with the same nuclear potentials as for $^6$He is about 5% below the experimental value $B_{GT}^{(\exp)}$ = 4.745 $\lambda^2$.

The cancellation explaining the small branching ratio appears to be essentially accidental. The delicate balance between the internal and external contributions is unlikely to occur in the beta decay of other halo nuclei. We have performed a similar calculation for the $^{11}$Li two-neutron halo nucleus [25]. This is made more difficult by the fact that the $^9$Li – n interaction is not known and other channels are open. The obtained branching ratio has an order of magnitude of $10^{-4}$. Although the agreement with an experiment performed after our calculation was published is qualitative only, the order of magnitude of the branching ratio is very close [26]. Cancellation effects are thus much weaker in this case.

## 6. M1 transition probabilities from the $^6$Li($0^+$) state

In Fig. 5, in order to check the energy dependence of the M1-transition, we display the differential width $d\Gamma_\gamma/dE = \hbar\, dW/dE$ for several $\alpha$+d potentials [27]. Contributions from three-body components up to $K_{max}$=20 are taken into account with the maximal relative distance $R_{max}$=30 fm. The folding potential $V_{f1}$ shows a picture strongly different from the other ones, with even a sharp minimum at about $E$ = 0.8 MeV. This potential gives a poor description of the $\alpha$ + d phase shift and hence a shifted node position for the $\alpha$ + d scattering wave function. This results in a strong cancellation effect as explained in the previous section. The folding potential $V_{f\,2}$ and the deep potential $V_a$ give close results and the supersymmetric potential $V_a^{S1}$ slightly overestimates them. The Gaussian potential $V_a$ simultaneously reproduces both the $^6$Li ground state binding energy and the $S$-wave phase shift at low energies. Additionally, the $S$-wave scattering wave function of this potential has two nodes at short distances (one due to the ground state, and one due to the Pauli forbidden state). The nearly phase-equivalent potential $V_{f2}$, which also has a forbidden bound state (and hence two nodes at short distances) gives similar results.

The influence of the nodes in the scattering wave function can be tested by using potential $V_a^S$. The corresponding width of the M1 transition is about two times larger. In the $^6$He β-decay process, this potential strongly overestimates the data [4]. Notice that a very different result is obtained with the folding potential $V_{f1}$, which has two bound states, but does not reproduce the $\alpha$ + d phase shifts and the $^6$He β decay. The shape and magnitude of the transition width and probability are strongly different from the result for $V_a$.

Considering the $V_a$ and $V_{f2}$ potentials, which are consistent with the data on $^6$He β decay, we deduce a recommended branching ratio of $1.3\times 10^{-4}$ by averaging both values. A previous estimate [13] of the branching ratio $\Gamma_\gamma$ $(0^+ \to \alpha + \mathrm{d})/\Gamma_\gamma(0^+ \to 1^+)$ provides $0.8\times 10^{-4}$. This value is close to our results obtained with potential $V_a$ [13]. Such a branching ratio should be observable experimentally.

## 7. Conclusions

In the present work, we studied the β-decay process of the $^6$He halo nucleus and the M1 transition of the $^6$Li($0^+$) analog state into the $\alpha$ + d continuum in the framework of a three-body model. Three-body hyperspherical bound-state wave functions on a Lagrange mesh and two-body $\alpha$ + d scattering wave functions have been used. For the calculation of the transition probabilities per time and energy units of the β decay, several $\alpha$ + d potentials were tested: an attractive Gaussian potential with a deep forbidden bound state, folding potentials derived from the α + $N$ $P$-wave potential and a repulsive potential [8].

The low experimental values result from a strong cancellation in the Gamow-Teller matrix element describing the transition to the continuum. This cancellation occurs between the internal and halo parts of the matrix element [9] and is thus very sensitive to the halo description. Reaching convergence is not easy: the two-body and three-body wave functions must extend up to 30-35 fm. From the analysis of the theoretical results we have found that converged results require the large value $K_{max}$=24 of the maximal hypermomentum. The dominant contributions to the transition probability come from $K$=2, $K$=8, and $K$=10 components of the three-body wave function. The contribution of the $K$=0 component is small due to an almost perfect cancellation of the internal and external parts of the Gamow-Teller matrix element.

The experimental transition probabilities per time and energy units [6] are well described with the slightly modified Gaussian potential which fairly reproduces the $^6$Li binding energy and the *S*-wave $\alpha$ + d phase shifts. The quality of the agreement arises from the node structure of the initial and final wave functions in the internal part. With the help of phase-equivalent potentials derived with supersymmetric transformations, we have shown that the role of the forbidden state is also essential. We realize that the efficiency of the deep potential may be somewhat fortuitous but the existence of a good agreement with experiment shows which ingredients are crucial in the interpretation of the β delayed deuteron decay of $^6$He.

The cancellation observed in the decay of $^6$He appears to be accidental and should probably not occur for other halo nuclei. This is confirmed by a similar calculation for $^{11}$Li [25] where the branching ratio is significantly larger in qualitative agreement with experiment [26].

In the M1 transition probability, the $K$ = 0 and $K$ = 2 components of the three-body wave function provide about 50% of the matrix elements; consequently, higher hypermomenta play an important role. The same conclusion holds in the $^6$He β decay into the $\alpha$ + d continuum, where large $K$ values cannot be neglected.

M1 transitions to the continuum provide a good probe of the halo structure in the $^6$Li($0^+$) state. The comparison with the $^6$He β decay shows that the inner parts of the matrix elements are very close to each other, as expected from charge symmetry. However, the halo parts are different, owing to the different binding energies, and different charges of the halo nucleons. In $^6$Li, the binding energy is lower, and therefore the asymptotic decrease of the wave function is slower. Consequently the halo contribution is larger in the γ-decay matrix element, and even represents the dominant part. This leads to the conclusion that charge-symmetry breaking is rather strong in these processes. The nearly perfect cancellation effect between short-range and halo contributions observed in $^6$He β-decay is less important here, and the sensitivity with respect to the potential is therefore weaker. Several α + d potentials were tested. The sensitivity is still important, but lower than in the $^6$He β-delayed decay. The present branching ratio of about $1.3\times 10^{-4}$ is consistent with the value of Ref. [13], where the authors use a simplified model. The present value is based on potential $V_a$ which reproduces the $^6$Li binding energy, the α + d low-energy phase shifts, and provides good results for the $^6$He β decay. It is therefore expected to have the same quality for the $^6$Li γ-decay. An experimental measurement seems to be possible with current facilities, and would provide, in combination with the data on $^6$He β decay, an important step in a better understanding of the halo structure in isobaric analog states.

---

1. I. Tanihata, J. Phys. **G 22**, 157 (1996).

2. P.G. Hansen and B. Jonson, Europhys. Lett. **4**, 409 (1987).

3. M.V. Zhukov, D.V. Fedorov, B.V. Danilin, J.S. Vaagen, and J.M. Bang, Nucl. Phys. **A529**, 53 (1991).

4. K. Riisager, M.J.G. Borge, H. Gabelmann, P.G. Hansen, L. Johannsen, B. Jonson, W. Kurcewicz, G. Nyman, A. Richter, O. Tengblad, and K. Wilhelmsen, Phys. Lett. **B 235**, 30 (1990).

5. M.J.G. Borge, L. Johannsen, B. Jonson, T. Nilsson, G. Nyman, K. Riisager, O. Tengblad, and K. Wilhelmsen Rolander, Nucl. Phys. **A560**, 664 (1993).

6. D. Anthony, L. Buchmann, P. Bergbusch, J.M. D'Auria, M. Dombsky, U. Giesen, K.P. Jackson, J.D. King, J. Powell, and F.C. Barker, Phys. Rev. **C 65**, 034310 (2002).

7. P. Descouvemont and C. Leclercq-Willain, J. Phys. G: Nucl. Part. Phys. **18**, L99 (1992).

8. M.V. Zhukov, B.V. Danilin, L.V. Grigorenko, and N.B. Shulgina, Phys. Rev. **C 47**, 2937 (1993).

9. D. Baye, Y. Suzuki, and P. Descouvemont, Prog. Theor. Phys. **91**, 271 (1994).

10. K. Varga, Y. Suzuki, and Y. Ohbayashi, Phys. Rev. **C 50**, 189 (1994).

11. F.C. Barker, Phys. Lett. **B322**, 17 (1994).

12. A. Csótó and D. Baye, Phys. Rev. **C 49**, 818 (1994).

13. L.V. Grigorenko and N.B. Shulgina, Phys. Atom. Nucl., **61**, 1472 (1998)

14. P. Descouvemont, C. Daniel, and D. Baye, Phys. Rev. **C 67**, 044309 (2003).

15. D. Baye and P.-H. Heenen, J. Phys. **A 19**, 2041 (1986).

16. S.B. Dubovichenko and A.V. Dzhazairov-Kakhramanov, Phys. Atom. Nucl. **57**, 733 (1994).

17. V.T. Voronchev, V.I. Kukulin, V.N. Pomerantsev, and G.G. Ryzhikh, Few-Body Syst. **18**, 191 (1995).

18. D.H. Wilkinson, Nucl. Phys. **A377**, 474 (1982).

19. E.M. Tursunov, D. Baye, and P. Descouvemont, Phys. Rev. **C 73,** 014303 (2006); **C 74**, 069904 (2006).

20. D.R. Thompson, M. LeMere, and Y.C. Tang, Nucl. Phys. **A268**, 53 (1977); I. Reichstein and Y.C. Tang, ibid. **A158**, 529 (1970).

21. H. Kanada, T. Kaneko, S. Nagata, and M. Nomoto, Prog. Theor. Phys. **61**, 1327 (1979).

22. W. Grüebler, P.A. Schmelzbach, V. König, R. Risler, and D. Boerma, Nucl. Phys. **A242**, 265 (1975).

23. B. Jenny, W. Grüebler, V. König, P.A. Schmelzbach, and C. Schweizer, Nucl. Phys. **A397**,

61 (1983).

24. D. Baye, Phys. Rev. Lett. **58**, 2738 (1987).

25. D. Baye, E.M. Tursunov, and P. Descouvemont, Phys. Rev. **C 74,** 064302 (2007)

26. R. Raabe, A. Andreyev, M. J. Borge, L. Buchmann, P. Capel, H. O. Fynbo, M. Huyse, R. Kanungo, T. Kirchner, C. Mattoon, A. C. Morton, I. Mukha, J. Pearson, J. Ponsaers, J. J. Ressler, K. Riisager, C. Ruiz, G. Ruprecht, F. Sarazin, O. Tengblad, P. Van Duppen, and P. Walden, Phys. Rev. Lett. **101**, 212501 (2008).

27. E.M. Tursunov, D. Baye, and P. Descouvemont, Nucl. Phys. **A793**, 2738 (2007).

TABLE I: Total transition probability per second $W$ (in $10^{-6}$ $s^{-1}$) for the beta decay of $^6$He into the $\alpha$ + d continuum.

| | $V_m$ | $V_{f1}$ | $V_{f2}$ | $V_r$ | Exp.[5] | Exp.[6] |
|---|---|---|---|---|---|---|
| $E$ > 0 MeV | 2.04 | 18.52 | 4.28 | 251 | | $2.2 \pm 1.1$ |
| $E$>0.37MeV | 1.59 | 13.43 | 2.45 | 182 | $7.6 \pm 0.6$ | $1.5 \pm 0.8$ |

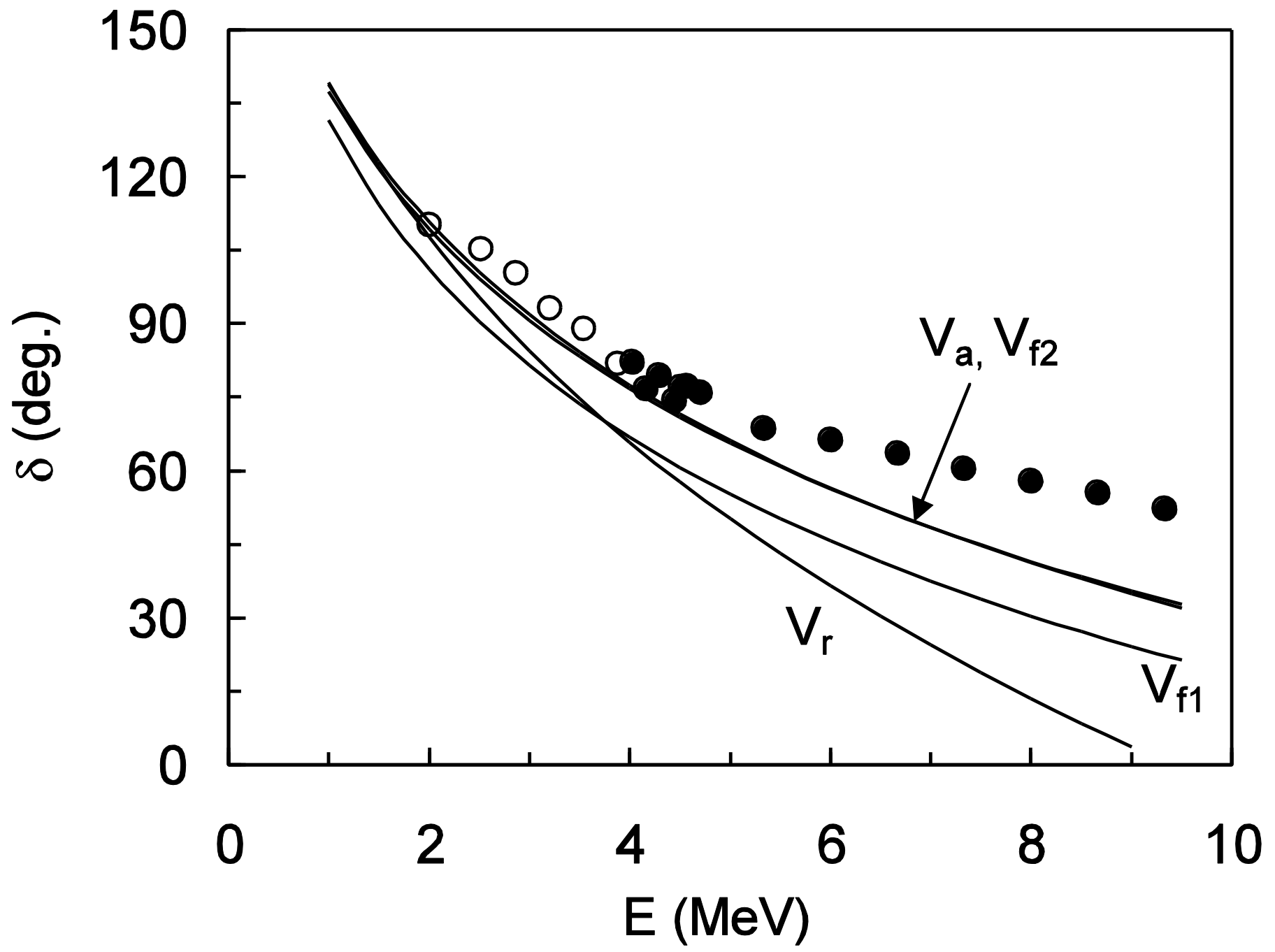

**Fig.1:** *S*-wave phase shifts obtained with different $\alpha$ +d potentials: attractive $V_a$ [16], folding potentials $V_{f1}$ and $V_{f2}$, and repulsive Wood-Saxon potential $V_r$. Phase-shifts are taken from analyses of experimental data in Refs. [22] (open dots) and [23] (full dots).

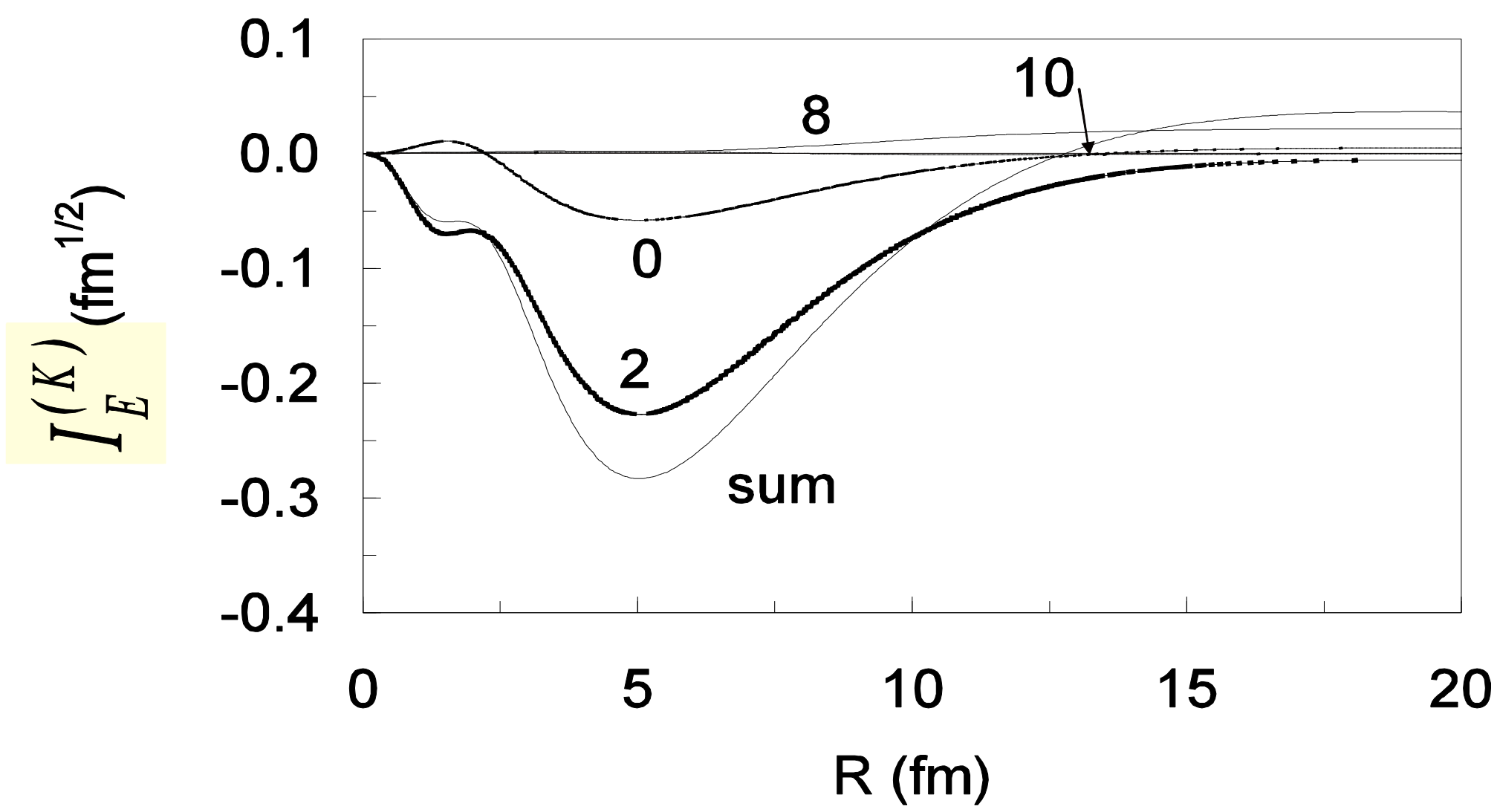

**Fig.2:** Effective integrals $I_E^{(K)}(R)$ at the energy $E$ = 1 MeV for the $\alpha$ +d potential of Ref. [16] and different $K$-values with $K_{max}$=24 and $R_{max}$=30 fm.

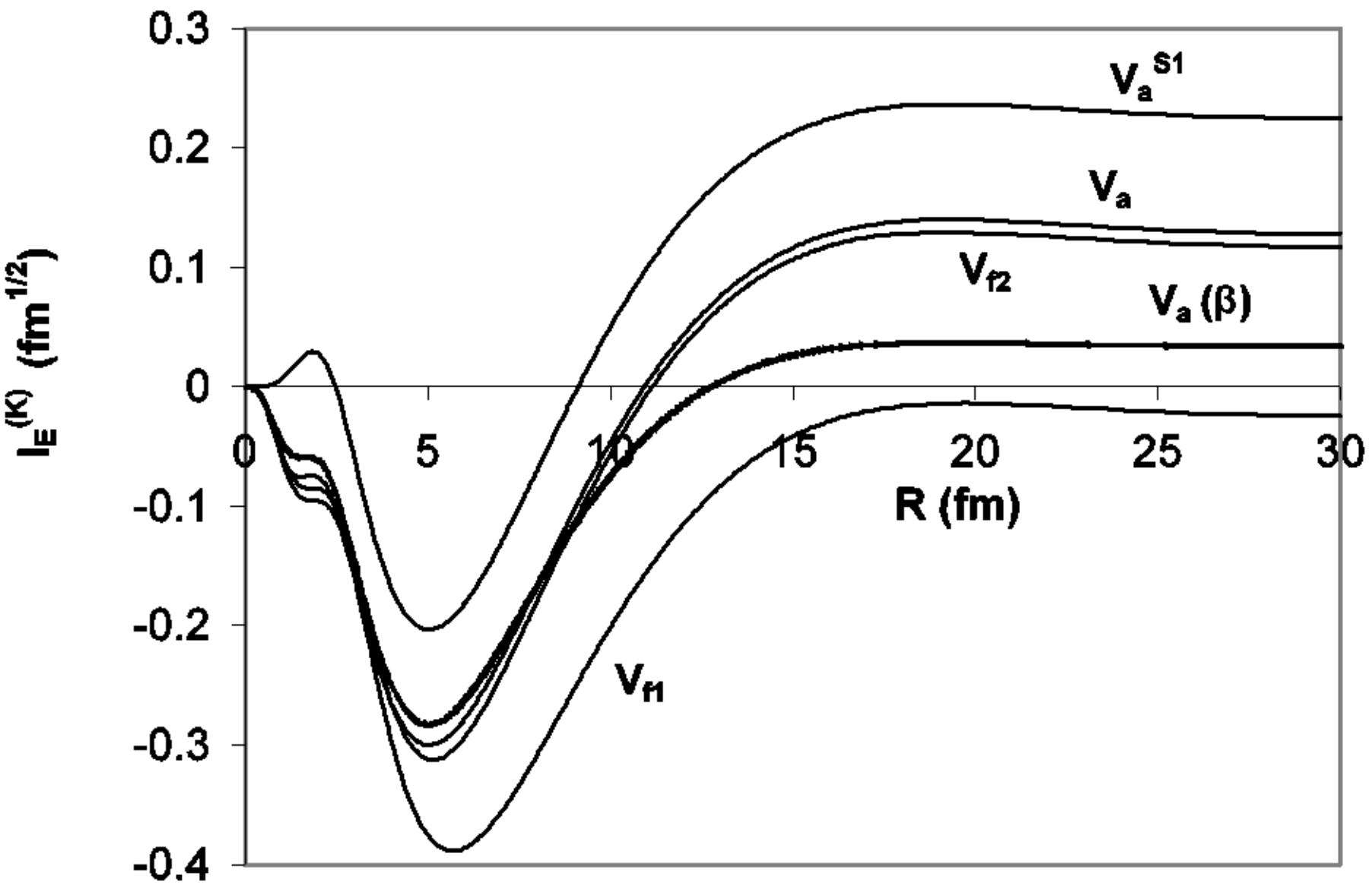

**Fig.3:** Effective integrals $I_E(R)$ at the energy $E$ = 1 MeV for different $\alpha$ +d potentials. $V_\alpha$(β) means the effective integral for the beta-decay.

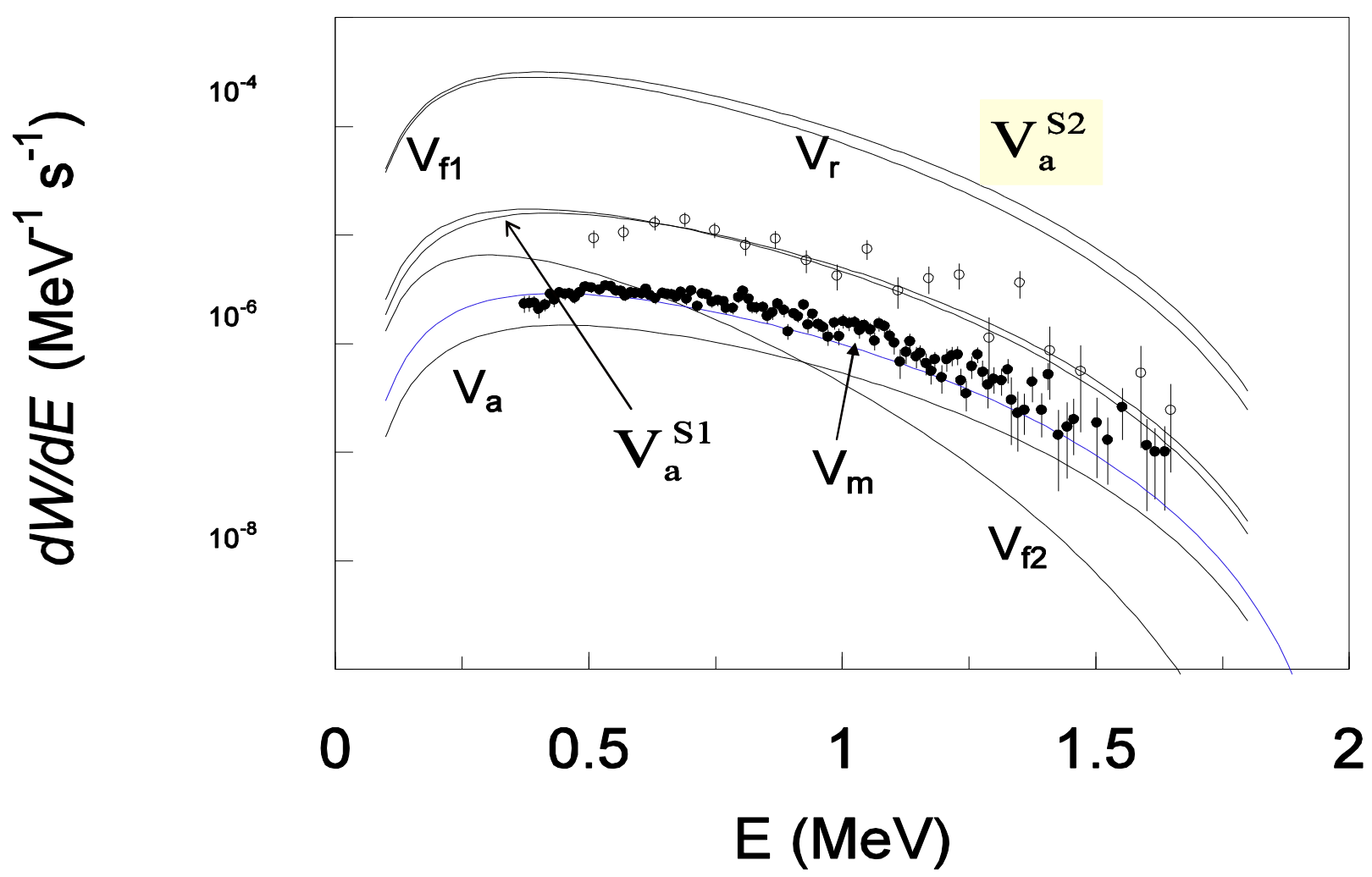

**Fig.4:** Transition probability per time and energy units *dW/dE* of the $^6$He β -decay into the $\alpha$ + d continuum with different $\alpha$ + d potentials for $K_{max}$=24 and $R_{max}$=30 fm. The experimental data *Exp.1* and *Exp.2* are from Ref. [5] and [6], respectively.

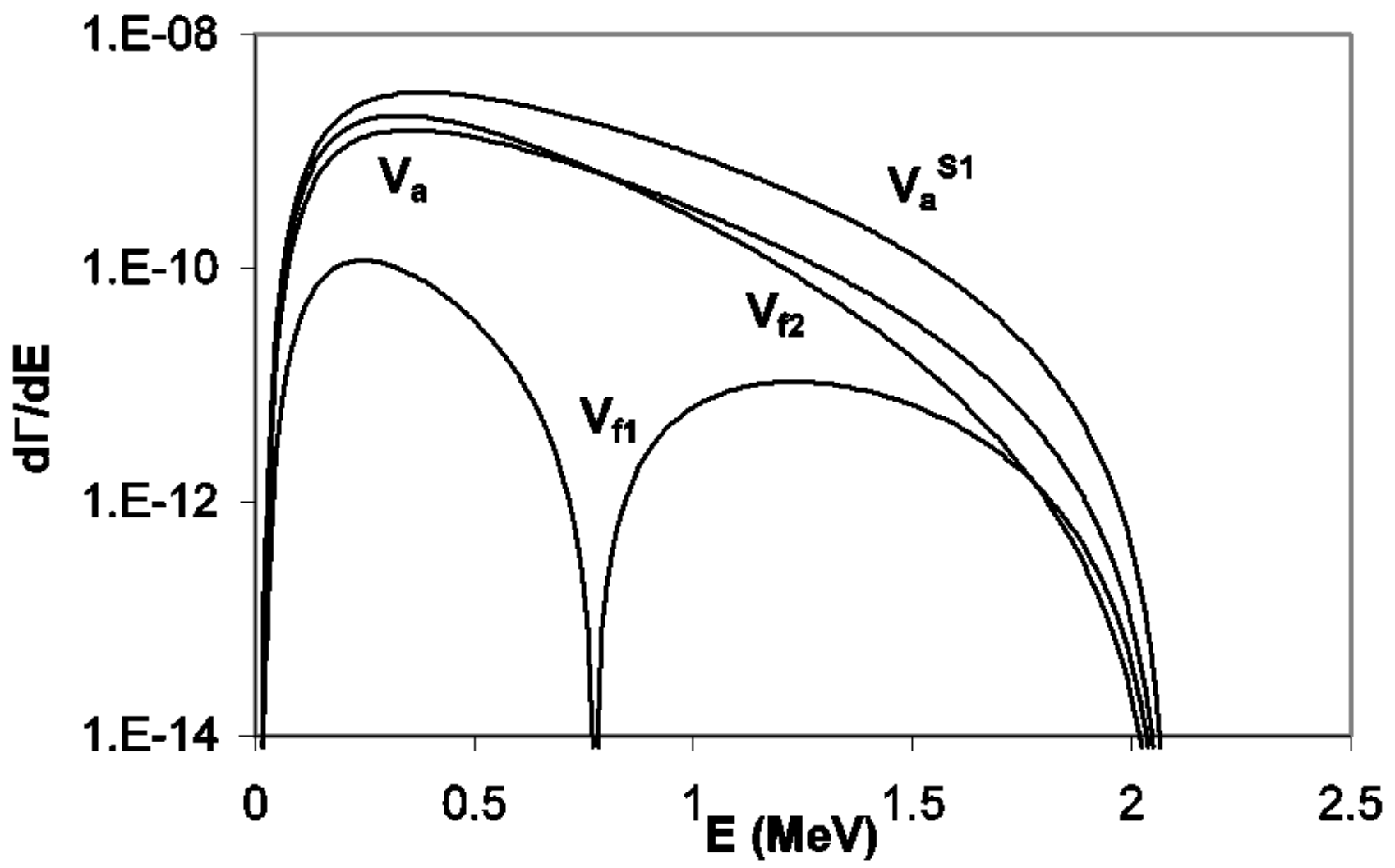

**Fig.5:** Relative width for the M1 transition process into the $\alpha$ +d continuum for several $\alpha$ + d potentials.